\documentclass[preprint,aps,prd,showpacs,nofootinbib,superscriptaddress,floatfix]{revtex4-1}
\usepackage{amsmath}
\usepackage{graphicx}
\usepackage{epstopdf}
\usepackage{epsfig}
\usepackage{amssymb}
\usepackage{bm}
\usepackage{braket}
\usepackage{comment}

\newcommand{\beq}{\begin{equation}}
\newcommand{\eeq}{\end{equation}}

\usepackage{color}

\allowdisplaybreaks[1]
\begin{document}

\begin{flushright}
%{\small KIAS-PREPRINT-******} 
\end{flushright}

\title{Stability of non-topological string \\
in supersymmetric $SU(2)\times U(1)$
gauge theory}

\author{Yukihiro Kanda}
\email[]{kanda.y@eken.phys.nagoya-u.ac.jp}
\affiliation{Department of Physics,
Nagoya University, Nagoya 464-8602, Japan}

\author{Nobuhiro Maekawa}
\email[]{maekawa@eken.phys.nagoya-u.ac.jp}
\affiliation{Department of Physics,
Nagoya University, Nagoya 464-8602, Japan}
\affiliation{Kobayashi-Maskawa Institute for the Origin of Particles and the
Universe, Nagoya University, Nagoya 464-8602, Japan}

\date{\today}

%\begin{abstract}
%\noindent
%\end{abstract}

\begin{abstract}
\noindent
We construct a non-topological string solution for a supersymmetric gauge theory with $SU(2)\times U(1)$
gauge symmetry which is spontaneously broken to $U(1)$ by developing the vacuum
expectation value of two doublet Higgses. It is a supersymmetric extension of the electroweak
string while supersymmetry is unbroken. We discuss the classical stability of the
non-topological string by perturbations.
%parameter region in which
%the string becomes stable classically, although it is unstable quantumly.
We show that the classical stability are determined only by two parameters, and that the allowed region becomes essentially the same as in the electroweak string. 
\end{abstract}

%\pacs{}

\maketitle

%%%%%%%%%%%%%%%%%%%%%%%%%%%%%%%%%%%%%%%%%%%%%%%%%%%%%%%
\section{Introduction}
\label{sec:intro}
Since the gravitational waves from black hole mergers were observed at LIGO\cite{LIGOScientific:2016aoc}, 
examining various models on elementary particle physics by primordial gravitational waves have been actively discussed\cite{Huang:2016cjm, Schwaller:2015tja,Chakrabortty:2020otp, Lazarides:2019xai}. One of them is to obtain evidence of symmetry breaking by
observing gravitational waves from cosmic strings\cite{Chakrabortty:2020otp, Lazarides:2019xai}, which are produced during symmetry
breaking\cite{Kibble:1976sj, Hindmarsh:1994re}.  
Recently, NANOGrav experiment reported the candidate of gravitational waves\cite{NANOGrav:2020bcs},
which is consistent with the signal of gravitational waves from cosmic strings\cite{Ellis:2020ena, Buchmuller:2020lbh, Blasi:2020mfx}.
This signal suggests the symmetry breaking scale is aound $10^{14-16}$ GeV\cite{Blasi:2020mfx, Chigusa:2020rks}.
\footnote[1]{The tension between NANOGrav and PPTA observation\cite{Shannon:2015ect} is known in the disscussion of gravitational waves from simple topological string. Several ideas to avoid this tension have been discussed\cite{Buchmuller:2020lbh, Buchmuller:2021dtt}.}
Therefore, cosmic strings research becomes increasingly important.

However,  most of studies on cosmic strings are concerned with topological 
strings, such as the Nielsen-Olesen strings\cite{Nielsen:1973cs} and $\mathbb{Z}_2$ strings\cite{Kibble:1982ae}, whose stability is guaranteed by topology of the vacuum and few are concerned with non-topological strings\cite{Achucarro:1999it, Eto:2021dca} which are quantumly unstable. One of the most commonly studied 
non-topological strings is the electroweak string\cite{Achucarro:1999it}, which had been discussed to be produced in breaking
the electroweak symmetry $SU(2)_L\times U(1)_Y$ in the standard model (SM). The idea of the electroweak string has been proposed in Ref. \cite{Nambu:1977ag} by Nambu, and the realistic solution as a cosmic string in the electroweak phase transition has been shown in Ref. \cite{Vachaspati:1992fi} by Vachaspati. The classical stability is determined only by two parameters, the Weinberg angle $\sin\theta_W$ and the ratio of Higgs mass to $Z$ boson mass $\beta\equiv m_H^2/m_Z^2$, and the parameter region,
in which the electroweak string becomes classically stable, has been numerically calculated
in Ref. \cite{James:1992zp, James:1992wb}.  Unfortunately, the electroweak
string becomes unstable even classically with the measured parameters in the SM.
However, this is not the case for the models beyond the SM, in which the parameters
have not been measured yet.

In this paper, we examine a non-topological string in the supersymmetric (SUSY) 
$SU(2)\times U(1)$ gauge theory. This is nothing but a SUSY extension of the electroweak string,
while the SUSY breaking is  neglected because we consider the physics at much higher scale than the electroweak scale such as the grand unified theories with $SO(10)$ \cite{Fritzsch:1974nn, Georgi:1974my, Clark:1982ai, Aulakh:1982sw} or $E6$ \cite{Gursey:1975ki, Achiman:1978vg, Shafi:1978gg, Mohapatra:1986bd}
unified group. We just embed the electroweak string configuration
into the SUSY $SU(2)\times U(1)$ gauge theory which has two doublet Higgs multiplets.
We will show that the stability of this non-topological string can be determined only
by two parameters, which are essentially the same as in the electroweak string.  
It is obvious that  the dangerous mode, which determine the allowed parameter region 
in the electroweak string, 
exists also in this SUSY extension of the electroweak string. Therefore, there is no possibility
to expand the  allowed parameter region. The question is whether the allowed region becomes smaller or not.  We will show that the allowed region of this non-topological 
string is the same as that of the electroweak string. Actually, we will show that all other 
modes do not destabilize the non-topological string. This is the main result of this paper.
  
In section 2, we review the electroweak string which can be constructed by embedding
the Nielsen-Olesen string into the $SU(2)\times U(1)$ gauge theory. The classical stability
conditions are reminded. In section 3, we construct the non-topological string 
by imposing the similar ansatz to the electroweak string on the SUSY $SU(2)\times U(1)$ gauge theory with two doublet Higgs multiplets. In section 4, we study the classical stability conditions of the non-topological string. 
The summary and discussion are devoted in section 5.

\section{Review of electroweak string}
First, we will review the electroweak string since we will discuss the SUSY extension
of it. The electroweak string can be constructed in a gauge theory
with $SU(2)_L\times U(1)_Y$ gauge symmetry which is spontaneously broken by developing the vacuum expectation value (VEV) of a doublet Higgs. This theory is similar to the standard model.

\subsection{Brief review of Nielsen-Olesen string}
\label{Nielsen-Olesen string solution}

First of all, we review the Nielsen-Olesen string\cite{Nielsen:1973cs} which is constructed in an $U(1)$ gauge theory
because the electroweak string can be constructed by embedding the Nielsen-Olesen string
into $SU(2)_L\times U(1)_Y$ gauge theory.

Let us consider an $U(1)$ gauge theory with a charged complex scalar $\phi(x)$ whose
Lagrangian $\mathcal{L}$ is given as 
\begin{align}
    \label{u1lagrangian}
    \mathcal{L} = -\frac{1}{4}F_{\mu\nu}F^{\mu\nu} + (D_\mu \phi)^* D^\mu\phi - V(\phi) ,
\end{align}
where $F_{\mu\nu}$ is the field strength of the $U(1)$ gauge field $A_\mu(x)$ and 
\begin{align}
    D_\mu \phi \equiv \partial_\mu \phi - i e A_\mu \phi.
\end{align}
Here, $e$ is the $U(1)$ gauge coupling constant and the potential $V(\phi)$ is written by
\begin{align}
    V(\phi) = \lambda \left( \phi^*\phi - \frac{v^2}{2}\right)^2 \qquad (\lambda>0, v>0).
\end{align}
%the moduli space of the VEV of $\phi(x)$ becomes 
%の真空期待値$\braket{\phi}$がとり得る値の集合である真空空間$\mathcal{V}$は、
%\begin{align}
 %   \mathcal{V} = \left\{ \phi \, | \, |\phi|^2 = \frac{v^2}{2} \right\}.
%\end{align}
Without loss of generality, the global minimum of the potential is given as
%となる。$\phi(x), A_\mu(x)$の古典解として、例えば次のような$\phi_v(x), A_{v\mu}(x)$が考えられる。
\begin{align}
    \phi_v(x) = \frac{v}{\sqrt{2}}, 
\end{align}
which becomes classical solution of the equations of motion with
\begin{align}
    A_{v\mu}(x) = 0.
\end{align}
These give the minimum value of the energy of this system.
%$\phi_v(x), A_{v\mu}(x)$の配位は系のエネルギーがglobal minimumになる配位でもある。$\phi_v(x), A_{v\mu}(x)$は静的な解であるため、時間依存しない場による系のエネルギー$E$
Actually, the energy for static solutions is given as
\begin{align}
    \label{tindpeu1}
    E = \int d^3x \left[ \frac{1}{4} F_{ij}F_{ij} + \left( D_i \phi \right)^* D_i \phi + V(\phi), \right]
\end{align}
and the above solution give the minimum energy $E=0$.

Let us explain  another stable and static classical solution with $E> 0$, which was
found by Nielsen and Olesen\cite{Nielsen:1973cs}.
The solutions with the translational symmetry in $z$ direction are 
%真の真空以外にも、$E \neq 0$であるが安定な古典解を作ることができる。静的かつ3次元空間の円筒座標$(r, \theta, z)$で$z$軸並進対称な次のansatz
\begin{align}
    \label{NOansatzp}
    \phi_s(x) &= f(r) e^{in\theta} \qquad (n\in\mathbb{Z}\backslash \{0\}) \\
    \label{NOansatzv}
    \vec{A}_s (x) &= \frac{n a(r)}{r} \vec{e}_\theta,
\end{align}
where  $f(r)$ and $a(r)$ are real valued functions of $r$  with the following
boundary conditions:
\begin{align}
    \label{nobc0}
    f(0) &= a(0) = 0 \\
    \label{nobcinf}
    f(\infty) &= \frac{v}{\sqrt{2}} , \qquad a(\infty) = \frac{1}{e}.
\end{align}
Here, we use cylindrical coordinates $(r, \theta, z)$ whose unit vectors are 
$(\vec{e}_r, \vec{e}_\theta, \vec{e}_z)$.
The equations of motion are given by
%を満たす動径$r$の実関数である。$\phi_s(x), \vec{A}_s(x)$が安定であるための必要条件として、系の古典解であることを要請する。運動方程式である(\ref{u1eomp})式、(\ref{u1eomg})式に$\phi_s(x), \vec{A}_s(x)$を代入すると、$f(r), a(r)$についての方程式が得られる。
\begin{align}
    \label{noeomf}
    &f''(r) + \frac{f'(r)}{r} - n^2 \frac{f(r)}{r^2} \left( 1 - ea(r) \right)^2 + 2 \lambda \left( \frac{v^2}{2} - f^2(r) \right) f(r) = 0, \\
    \label{noeomv}
    &a''(r) - \frac{a'(r)}{r} + 2 e n f^2(r) \left( 1 - ea(r) \right) = 0,
\end{align}
where $f'(r) \equiv \frac{d}{dr} f(r)$. Note that satisfying the equations of motion is a necessary condition to obtain stable solutions. These are the Nielsen-Olesen string solutions.
%である。この方程式を満たす$f(r), a(r)$での(\ref{NOansatzp})式、(\ref{NOansatzv})式がNielsen-Olesen string解である。

It is guaranteed by topological feature of the moduli space that the Nielsen-Olesen string 
solutions  do not decay to the global minimum. Generally, if the first homotopy group to 
the moduli of the breaking $G\rightarrow H$, $\pi_1(G/H)$, is non-trivial, a stable string solution appears. The first homotopy group of the breaking $U(1)\rightarrow \times$ 
becomes non-trivial as $\pi_1(U(1))=\mathbb{Z}$. 
A string whose stability is guaranteed by topology is called a topological string, while
the stability of a non-topological string is not guaranteed by topology, i.e., the first homotopy
group to the moduli becomes trivial as $\pi_1(G/H)=1$.

\subsection{Electroweak string}
One of the most interesting examples of non-topological strings is the electroweak string, which may appear in breaking $SU(2)\times U(1)\rightarrow U(1)$ by a doublet Higgs, 
although any topological strings cannot appear, i.e., $\pi_1(SU(2)\times U(1)/U(1))=1$.
The stability of the electroweak string has been discussed in Ref. \cite{James:1992zp, James:1992wb}, and unfortunately, for the measured parameters in the SM, 
the electroweak string becomes unstable.

However, this may not the case for the models beyond the standard model. Therefore,
the produced string produces gravitational waves which may be observed in near future.

In this subsection, we will review the electroweak string  briefly.
First, let us consider the $SU(2)_L\times U(1)_Y$ gauge theory with a doublet Higgs with 
$U(1)_Y$ charge $1/2$, whose Lagrangian  is given as
\begin{align}
    \label{ewtranslagrangian}
    \mathcal{L} = -\frac{1}{4}W^a_{\mu\nu}W^{a\mu\nu} -\frac{1}{4}F_{\mu\nu}F^{\mu\nu} + (D_\mu H)^\dagger D^\mu H + \mu^2 H^\dagger H - \lambda (H^\dagger H)^2,
\end{align}
where $W_{\mu\nu}^a$ and $F_{\mu\nu}$ are the field strengths of $SU(2)_L$ and 
$U(1)_Y$, respectively. 
The moduli space of $H$ becomes
\begin{align}
    \label{ewvacspace1}
    \mathcal{V}_{EW} = \left\{ H \middle| H^\dagger H = \frac{\mu^2}{2 \lambda}
\equiv \frac{v^2}{2}\right\},
\end{align}
where $v>0$. Since $\pi_1(\mathcal{V}_{EW})$ is trivial, a topological string does not appear
in this model.

When the VEV of $H$ becomes
\begin{align}
    H = \left(
        \begin{array}{c}
            0 \\
            \frac{v}{\sqrt{2}}
        \end{array}
    \right),
\end{align}
the kinetic term of the Higgs field gives the mass terms of the gauge fields as
\begin{align}
    (D_\mu H)^\dagger D^\mu H &\supset
    \left(
        \begin{array}{cc}
            0 & \frac{v}{\sqrt{2}}
        \end{array}
    \right)
    \left( g W^a_\mu \frac{\sigma^a}{2} + g' B_\mu \frac{1}{2} \right)^2 
    \left(
        \begin{array}{c}
            0 \\ \frac{v}{\sqrt{2}}
        \end{array}
    \right) \nonumber\\
    &= \frac{v^2}{2} \left[ \frac{g^2}{4} W^{a}_\mu W^{a\mu} -\frac{gg'}{2} W^3_\mu B^\mu + \frac{g'^2}{4} B_\mu B^\mu \right] \nonumber\\
    \label{wzmassterm}
    &= \frac{g^2v^2}{8} W^{\bar{a}}_\mu W^{\bar{a}\mu} +\frac{\alpha^2v^2}{8} \left( \frac{g}{\alpha} W^3_\mu - \frac{g'}{\alpha} B_\mu \right) \left( \frac{g}{\alpha} W^{3\mu} - \frac{g'}{\alpha} B^\mu \right) \qquad (\bar{a} = 1,2),
\end{align}
where $g$ and $g'$ are the gauge coupling constants of $SU(2)_L$ and $U(1)_Y$, 
respectively, $\alpha\equiv \sqrt{g^2+g'^2}$, and $\sigma^a/2$ are 
generators of $SU(2)_L$.
The $W$ boson and the $Z$ boson as 
$Z\equiv \cos\theta_W W^3_\mu - \sin\theta_W B_\mu$ obtain the masses 
$gv/2$ and $\alpha v/2$, respectively.
Here $\cos\theta_W=g/\alpha$. 
On the other hand, the photon,
$A_\mu = \sin\theta_W W^3_\mu + \cos\theta_W B_\mu$, remains massless, which
is the gauge field corresponding to unbroken $U(1)_{EM}$.

Let us consider $U(1)_L$ whose generator is $\sigma_3/2$, instead of $SU(2)_L$. 
The VEV of Higgs breaks $U(1)_L\times U(1)_Y=U(1)_Z\times U(1)_{EM}$ into $U(1)_{EM}$.
Since the first homotopy group for this breaking is non-trivial, a topological string appears.
This is nothing but the Nielsen-Olesen string for $U(1)_Z$ gauge theory with gauge coupling
constant $-\alpha/2$.  When this Nielsen-Olesen string solution is embedded in the 
$SU(2)_L\times U(1)_Y$ gauge theory as
\begin{align}
    \label{ewstringconfig}
    \begin{aligned}
        &H(x) = \left(
        \begin{array}{c}
            0 \\ \phi_s(x)
        \end{array}
        \right) , \qquad Z_\mu(x) = A_{s\mu}(x) \\
        &A_\mu(x) = W^{\bar{a}}_\mu(x) = 0,
    \end{aligned}
\end{align}
these satisfy the equations of motion. This is the electroweak string solution.

\subsection{Stability of electroweak string}
\label{stabilityofEWstringsec}
Although the electroweak string becomes a classical solution, it is not obvious that this 
solution is classically stable or not. The stability can be determined by the presence or absence of perturbation modes which make the energy lower.

Let us consider the perturbations from the $n=1$ electroweak string
\begin{align}
    \label{ewstringansatz}
    \begin{aligned}
        &H(x) = \left(
        \begin{array}{c}
            0 \\ f(r) e^{i\theta}
        \end{array}
        \right) , \qquad Z_0(x) = 0, \quad \vec{Z}(x) = -\frac{z(r)}{r} \vec{e}_\theta \\
        &A_\mu(x) = W^{\bar{a}}_\mu(x) = 0
    \end{aligned}
\end{align}
as
\begin{align}
    \label{ewstringperturbation}
    \begin{aligned}
        &H(x) = \left(
        \begin{array}{c}
            h(x) \\ f(r) e^{i\theta} + \delta \phi(x)
        \end{array}
        \right) , \qquad Z_0(x) = \delta Z_0(x), \quad \vec{Z}(x) = -\frac{z(r)}{r} \vec{e}_\theta + \delta\vec{Z}(x) \\
        &A_\mu(x) = A_\mu(x), \qquad W^{\bar{a}}_\mu(x) = W^{\bar{a}}_\mu(x).
    \end{aligned}
\end{align}
Since the $t$ and $z$ dependence of the perturbations and non-vanishing $t$ and $z$ component gauge fields increases the energy, we take the perturbations independent of $t$ and $z$, and we ignore the $t$ and $z$ component gauge fields.
Thus, the energy linear density (string tension) $\mu$ becomes
\begin{align}
    \label{ewenergydensity}
    \mu = \int rdrd\theta \left[ \frac{1}{4} \left( W^a_{\bar{i}\bar{j}} \right)^2 + \frac{1}{4} \left( F_{\bar{i}\bar{j}} \right)^2 + \left( D_{\bar{i}} H \right)^\dagger D_{\bar{i}} H - \mu^2 H^\dagger H + \lambda \left( H^\dagger H \right)^2 \right] \qquad (\bar{i}, \bar{j} = 1, 2).
\end{align}
Substituting the electroweak string solution (\ref{ewstringansatz}) into (\ref{ewenergydensity}), the string tension becomes
\begin{align}
    \label{ewstringmu}
    \mu_0 = \int rdrd\theta \left[ \frac{z'}{2r^2} + f'^2 + \left( 1 - \frac{\alpha}{2}z \right)^2 \frac{f^2}{r^2} + \lambda \left( f^2 - \frac{v^2}{2} \right)^2 \right],
\end{align}
which is the tension of the electroweak string.
Since the electroweak string solution satisfies the classical equations of motion, the 
leading terms of the variation of the string tension are quadratic terms of these
perturbation modes $\delta\phi(x), \delta Z_\mu(x), h(x), A_\mu(x), W^{\bar{a}}_\mu(x)$.
Because of conservation of $U(1)_{EM}$ charge, the quadratic terms of neutral part 
$\delta\mu_{\delta\phi, \delta Z, A}$ and charged part $\delta\mu_{h.W}$ are separated,
i.e., $\delta\mu=\mu-\mu_0\sim \delta\mu_{\delta\phi,\delta Z, A}+\delta\mu_{h.W}$.
The neutral part $\delta\mu_{\delta\phi, \delta Z, A}$ must be non-negative because
this part is the same as the Nielsen-Olesen string which is stable. It has been also shown numerically in Ref. \cite{Goodband:1995he}. 
Therefore, we have to examine whether $\delta\mu_{h,W}$
\begin{align}
    \delta\mu_{h,W} &= \int rdrd\theta \left[ \left| \left( \partial_{\bar{i}} - i \frac{\alpha\cos 2\theta_W}{2} Z_{\bar{i}} \right) h \right|^2 + 2\lambda \left( f^2 - \frac{v^2}{2} \right) h^* h \right. \nonumber\\
    & \hspace{2cm} + i \frac{\alpha\cos\theta_W}{2} W^{\bar{a}}_{\bar{i}} \left( H^\dagger \sigma^{\bar{a}} d_{\bar{i}} H - \left(d_{\bar{i}} H\right)^\dagger \sigma^{\bar{a}} H \right) \nonumber \\
    & \hspace{2cm} + \frac{g^2}{4} f^2 \vec{W}^{\bar{a}} \cdot \vec{W}^{\bar{a}} + \alpha \cos^2\theta_W \left(\vec{W}^1 \times \vec{W}^2\right) \cdot \left( \nabla \times \left( -\frac{z(r)}{r} \vec{e}_\theta \right) \right) \nonumber\\
    & \hspace{2cm} + \frac{1}{2} \left| \nabla \times \vec{W}^1 + \alpha\cos^2\theta_W \vec{W}^2 \times \left( -\frac{z(r)}{r} \vec{e}_\theta \right) \right|^2 \nonumber\\
    & \hspace{2cm} \left. + \frac{1}{2} \left| \nabla \times \vec{W}^2 - \alpha\cos^2\theta_W \vec{W}^1 \times \left( -\frac{z(r)}{r} \vec{e}_\theta \right) \right|^2 \right]
\end{align}
can be negative or not.

The perturbation modes $h(x), \vec{W}^{\bar{a}}(x)$ can be expanded as
\begin{align}
    \label{ewhexp}
    h(x) &\equiv \sum_{m=-\infty}^\infty \chi_m(r) e^{im\theta} \\
    \vec{W}^1(x) &\equiv \sum_{m=0}^\infty \left[  \left( W^1_m(r) \sin m\theta + \overline{W}^1_m(r) \cos m\theta \right) \vec{e}_r + \frac{1}{r} \left( \omega^1_m(r) \cos m\theta + \overline{\omega}^1_m(r) \sin m\theta \right) \vec{e}_\theta \right] \\
    \vec{W}^2(x) &\equiv \sum_{m=0}^\infty \left[  \left( W^2_m(r) \cos m\theta + \overline{W}^2_m(r) \sin m\theta \right) \vec{e}_r + \frac{1}{r} \left( \omega^2_m(r) \sin m\theta + \overline{\omega}^2_m(r) \cos m\theta \right) \vec{e}_\theta \right].
\end{align}
Thus, the variation of string tension by these deformations becomes
\begin{align}
   \delta \mu_{h, W} &= 2\pi \int rdr \sum_m \left[ \varepsilon_{\chi, m} + \varepsilon_{c, m} + \varepsilon_{W, m} \right] \\
    \label{ewepsx}
    \varepsilon_{\chi, m} &= \left|{\chi'}_m\right|^2 + \left\{ \frac{1}{r^2}\left(m + \frac{\alpha\cos 2\theta_W}{2}z\right)^2 + 2\lambda\left(f^2 - \frac{v^2}{2}\right) \right\} \left|\chi_m\right|^2 \\
    \label{ewepsc}
    \varepsilon_{c, m} &= \frac{\alpha\cos\theta_W}{4} \nonumber\\
    &\hspace{0.5cm} \times\left\{ \left( \chi^*_mf' - {\chi_m^*}'f \right)\left(i\overline{W}^1_{|1-m|} + W^2_{|1-m|} + \left(-W^1_{|1-m|}+i\overline{W}^2_{|1-m|}\right)\mathrm{sgn}(1-m) \right) \right. \nonumber\\ 
    & \hspace{1cm}\left. +\left( \chi_mf' - {\chi_m}'f \right) \left( -i\overline{W}^1_{|m-1|} + w^2_{|m-1|} + \left(W^1_{|m-1|}+i\overline{W}^2_{|m-1|}\right)\mathrm{sgn}(m-1) \right) \right\} \nonumber\\
    &\hspace{0.3cm}+ \frac{\alpha\cos\theta_W}{4r^2}  \left(1+m-\alpha\sin^2\theta_Wz\right)f \nonumber\\
    &\hspace{2cm} \times\left\{ \chi^*_m\left( -\omega^1_{|1-m|} + i\overline{\omega}^2_{|1-m|} + \left(-i\overline{\omega}^1_{|1-m|} - \omega^2_{|1-m|}\right)\mathrm{sgn}(1-m) \right) \right. \nonumber\\
    &\left. \hspace{3cm}+ \chi_m \left( -\omega^1_{|m-1|} -i \overline{\omega}^2_{|m-1|} + \left( -i \overline{\omega}^1_{|m-1|} + \omega^2_{|m-1|} \right) \mathrm{sgn}(m-1) \right)\right\} \\
    \label{ewepsw}
    \varepsilon_{W, m} &= \frac{g^2f^2}{8} \left\{ \left( W^{\bar{a}}_m \right)^2 + \frac{1}{r^2} \left( \omega^{\bar{a}}_m \right)^2 \right\} + \frac{\alpha\cos^2\theta_W z'}{2r^2} \left( W^2_m \omega^1_m - W^1_m \omega^2_m \right) \nonumber\\
    &\hspace{0.3cm} + \frac{1}{4r^2} \left\{ \left( mW^1_m - {\omega^1_m}' + \alpha\cos^2\theta_WzW^2_m \right)^2 + \left( mW^2_m + {\omega^2_m}' + \alpha\cos^2\theta_W z W^1_m \right)^2 \right\} \nonumber\\
    &\hspace{0.3cm} + \left( W^1_m \rightarrow \overline{W}^1_m, W^2_m \rightarrow -\overline{W}^2_m, \omega^1_m \rightarrow -\overline{\omega}^1_m, \omega^2_m \rightarrow \overline{\omega}^2_m, \right).
\end{align}
In the above equations, for sets of modes 
$\chi_m, W^{\bar{a}}_{|m-1|}, \omega^{\bar{a}}_{|m-1|}, \overline{W}^{\bar{a}}_{|m-1|}, \overline{\omega}^{\bar{a}}_{|m-1|}$ $(m \in \mathbb{Z})$, the sets with different $m$ are independent
of each other. Among these sets, the set with $m=0$ must includes the most dangerous modes for stability because the main negative contribution from the second term in 
eq. (\ref{ewepsx}) becomes positive around $r\sim 0$ unless $m\neq 0$.
% negative around $r\sim 0$,  is dominated by the
% positive contribution from the first term around $r\sim 0$ unless $m=0$. 
Therefore, we fix $m=0$ in the
following discussion, and for simplicity, we take
$\chi \equiv \chi_0, W^{\bar{a}}\equiv W^{\bar{a}}_1, \omega^{\bar{a}}\equiv\omega^{\bar{a}}_1, \overline{W}^{\bar{a}}\equiv \overline{W}^{\bar{a}}_1, \overline{\omega}^{\bar{a}}\equiv\overline{\omega}^{\bar{a}}_1$.

Moreover, it is seen that 
\begin{align}
    \delta\mu_{h,W} \equiv \mu_{\mbox{even}}(\mathrm{Re}\, \chi, W^{\bar{a}}, \omega^{\bar{a}})
 + \mu_{\mbox{odd}}(\mathrm{Im}\, \chi, \overline{W}^{\bar{a}}, \overline{\omega}^{\bar{a}}),
\end{align}
where $\mathrm{Re}\, \chi, W^{\bar{a}}, \omega^{\bar{a}}$ are CP even modes while
$\mathrm{Im}\, \chi, \overline{W}^{\bar{a}}, \overline{\omega}^{\bar{a}}$ are CP odd modes.
Since $\mu_{\mbox{even}}$ is the same functional as $\mu_{\mbox{odd}}$, it is sufficient to
study one of the two functionals, for example, $\mu_{\mbox{even}}$, in order to check the
stability of this string solution.

$\mu_{\mbox{even}}$ is given as 
\begin{align}
    \mu_{\mbox{even}} &= 2\pi \int rdr \left[ {{\chi_R}'}^2 + \left\{ \frac{\left(1+\frac{\alpha}{2}\cos2\theta_Wz\right)^2}{r^2} + 2\lambda \left( f^2 - \frac{v^2}{2} \right) \right\} \chi_R^2 \right. \nonumber\\
    &\hspace{2cm} + \frac{\alpha\cos\theta_W}{2} \left(\chi_R f' - {\chi_R}' f\right) \left(W^2 - W^1\right) \nonumber\\
    &\hspace{2cm} - \frac{\alpha\cos\theta_W}{4r^2} \left(1-\alpha\sin^2\theta_Wz\right) f\chi_R \left(\omega^1 + \omega^2\right) \nonumber\\
    &\hspace{2cm} + \frac{g^2f^2}{8} \left\{ \left( W^{\bar{a}} \right)^2 + \frac{1}{r^2} \left( \omega^{\bar{a}} \right)^2 \right\} + \frac{\alpha\cos^2\theta_W z'}{2r^2} \left( W^2 \omega^1 - W^1 \omega^2 \right) \nonumber\\
    &\hspace{2cm}  + \frac{1}{4r^2} \left\{ \left( W^1 - {\omega^1}' + \alpha\cos^2\theta_W z W^2 \right)^2 \right. \nonumber\\
    &\hspace{4cm} \left. \left.+ \left( W^2 + {\omega^2}' + \alpha\cos^2\theta_W z W^1 \right)^2 \right\} \right].%\nonumber
\end{align}
where we write $\chi_R(r)$ instead of $\mathrm{Re}\, \chi$. 
By completing the square for variables ${W}^{\bar{a}}$,  we can neglect these variables by
taking the squares vanishing. As the result, we have only three variables, 
$\chi_R, \xi_\pm\equiv (\omega^2\pm \omega^1)/2$. And $\mu_{\mbox{even}}$ is again divided into two parts as
\begin{align}
    \mu_{\mbox{even}} &= \mu_{\xi_-} + \mu_{\chi, \xi_+}
\end{align}
\begin{align}
    \mu_{\xi_-} = \pi \int \frac{dr}{r} \left[ \frac{\alpha^2\cos^2\theta_W r^2 f^2}{2P_-} {\xi'}_-^2 + \left( \frac{\alpha^2\cos^2\theta_W f^2}{2} - \frac{\gamma^2\theta_W {z'}^2}{P_-} - r \frac{d}{dr} \left(\frac{\gamma z' (1+\gamma z)}{rP_-}\right) \right) \xi_-^2 \right]
\end{align}
\begin{align}
    \mu_{\chi, \xi_+} = &2\pi \int rdr \left[ \frac{\left(1-\gamma z\right)^2}{P_+} {{\chi_R}'}^2 \right. \nonumber\\
&+ \left\{ \frac{\left(1+\frac{\alpha}{2}\cos2\theta_Wz\right)^2}{r^2} + 2\lambda \left( f^2 - \frac{v^2}{2} \right) - \frac{\alpha^2\cos^2\theta_W r^2}{2P_+} \left( {f'}^2 - ff'\frac{d}{dr} \right) \right\} \chi_R^2 \nonumber\\
    & + \frac{\alpha^2\cos^2\theta_W r^2 f^2}{4 r^2 P_+} {\xi'}_+^2 + \left\{ \frac{\alpha^2\cos^2\theta_W f^2}{4r^2} - \frac{\gamma^2 {z'}^2}{2r^2P_+} - \frac{\gamma z' (1-\gamma z)}{2r^2P_+} \frac{d}{dr} \right\} \xi_+^2 \nonumber\\
    & \left. - \alpha\cos\theta_W \left\{ \frac{1-\alpha\sin^2\theta_W z}{r^2} f\chi_R\xi_+ + \frac{\chi_R f' - {\chi_R}' f}{P_+} \left( \left(1-\gamma z\right) {\xi'}_+ + \gamma z' \xi_+ \right) \right\} \right],
\end{align}
where $\gamma \equiv\alpha\cos^2\theta_W$ and 
\begin{align}
    P_\pm \equiv\left(1 \mp \alpha\cos^2\theta_W z\right)^2 + \frac{\alpha^2\cos^2\theta_W}{2} r^2 f^2.
\end{align}
It is confirmed by numerical calculation that $\mu_{\xi_-}$ does not include negative contribution\cite{James:1992wb}.
Since one of the linear combination of $\chi_R(r), \xi_+(r)$ is just the gauge transformation, 
the physical perturbation becomes
\begin{align}
    \zeta(r) = \left(1-\gamma z\right) \chi_R(r) + \frac{\alpha\cos\theta_W f}{2} \xi_+(r).
\end{align}
In summary, the classical stability of this electroweak string can be determined by the
following energy variation as
\begin{align}
    \label{ewmuzeta}
    &\mu_\zeta = 2\pi \int rdr \zeta \mathcal{O} \zeta \\
    \label{ewop}
    &\mathcal{O} \equiv -\frac{1}{r}\frac{d}{dr}\left(\frac{r}{P_+}\frac{d}{dr}\right) + \left\{ \frac{2S_+}{g^2r^2f^2} + \frac{{f'}^2}{f^2 P_+} + \frac{1}{r}\frac{d}{dr}\left(\frac{rf'}{fP_+}\right) \right\} \\
    &S_+ \equiv \frac{g^2f^2}{2} - \frac{\gamma^2{z'}^2}{P_+} + r\frac{d}{dr} \left\{ \frac{\gamma z' (1-\gamma z)}{r P_+} \right\}.
\end{align}
It is important to know whether the eigenfunction $\zeta(r)$ with negative eigenvalue for
 the operator $\mathcal{O}$. 
This numerical calculation has been done, and the allowed region in parameter space
$(\beta\equiv m_H^2/m_Z^2, \cos^2\theta_W)$ has been obtained in Ref. \cite{James:1992zp,James:1992wb}.

 \section{Non-topological string on supersymmetric $SU(2)\times U(1)$ model}

We consider the supersymmetric $SU(2)\times U(1)$ gauge theory  with the Higgs fields $H_1$ and $H_2$, whose quantum numbers are  $ (\bm{2}, \frac{1}{2})$ and $ (\bm{2}, -\frac{1}{2})$ under $SU(2)\times U(1)$, respectively,   and a gauge singlet $S$. The superpotential $W$ is
\begin{align}
    W = y S \left(H_2^t (i\sigma^2) H_1  - u^2 \right) ,
\end{align}
where $i \sigma^2$ is the $2\times 2$ antisymmetric matrix. We take parameters $y$ and $u$ real for simplicity. 
Although there are fermions as supersymmetric partner, we do not consider that
the fermions develop the non-vanishing VEVs. Thus, we consider the scalar components  $h_{1}$, $h_{2}$ and $s$ for chiral multiplets $H_{1}$, $H_{2}$ and $S$. 
The potential $V(h_1, h_2, s)$ becomes
\begin{align}
    \label{hpot}
    V(h_1, h_2, s) = &y^2 \left| h_2^t (i\sigma^2)h_1 - u^2 \right|^2 + y^2 |s|^2 
\left(|h_1|^2+|h_2|^2\right) \nonumber\\
    &+ \frac{g_2^2+g_1^2}{8} \left(|h_1|^2 - |h_2|^2\right)^2 + \frac{g_2^2}{2} \left| h_2^\dagger h_1 \right|^2 ,
\end{align}
where $g_{1}$ and $g_{2}$  are the gauge couplings of $U(1)$ and $SU(2)$, respectively.
  The Lagrangian for scalar and gauge fields is
\begin{align}
    \label{susyXL}
    \mathcal{L} &= -\frac{1}{4} W^a_{\mu\nu} W^{a\mu\nu} - \frac{1}{4} F_{\mu\nu} F^{\mu\nu} + \left| D_\mu h_1 \right|^2 + \left| D_\mu h_2 \right|^2 + \left| \partial_\mu s \right|^2 - V(h_1, h_2, s),
\end{align}
where the notation is the same as in section 2 for the $SU(2)\times U(1)$ gauge fields.
We take the same ansatz of solutions as in Ref. \cite{Earnshaw:1993yu}, in which the electroweak string in general two Higgs doublet models (2HDMs) are constructed
\cite{Earnshaw:1993yu}. 
\footnote[2]{Topological strings in 2HDMs has been actively researched\cite{Battye:2011jj}.}
The string solutions are
\begin{align}
    \label{susyewstringansatz}
    \begin{aligned}
        &h_1(x) = f_1(r) e^{im\theta} \left(\begin{array}{c} 0 \\ 1 \end{array}\right), & 
    &h_2(x) = f_2(r) e^{-im\theta} \left(\begin{array}{c} 1 \\ 0 \end{array}\right), &
    &\vec{Z}(x) = -m\frac{z(r)}{r} \vec{e}_\theta \\
    &(\mbox{other fields}) = 0 ,
    \end{aligned}
\end{align}
where $m$ is integer and non-zero. 
$f_1(r), f_2(r)$ and $z(r)$ are real-valued functions and satisfy the boundary conditions :
\begin{align}
    \label{susyewstringbc}
    &f_1(0) = f_2(0) = z(0) = 0 , \\
    &f_1(\infty) = f_2(\infty) = u, z(\infty) = \frac{2}{\alpha} .
\end{align}
Therefore we obtain the string solutions to be aligned the $z$ axis.
%With the eq. (\ref{susyewstringbc}), eq. (\ref{susyewstringansatz}) are the static string solutions to be aligned the $z$ axis.

We consider F- and D-flatness conditions\cite{Luty:1995sd} for eq. (\ref{susyewstringansatz}). First, F-flatness conditions give
\begin{align}
    \label{stringf-flat}
    \begin{aligned}
        &f_1(r) f_2(r) - u^2 = 0 , \\
        &s = 0 .
    \end{aligned}
\end{align}
With the boundary conditions given in (\ref{susyewstringbc}), $f_1(r)$ and $f_2(r)$ do not satisfy F-flatness condition. This indicates that supersymmetry is broken inside the string. 
%While it may be interesting to consider this effect for other phenomenology, we do not discuss them in this paper. 
On the outside the string, eq. (\ref{stringf-flat}) is almost satisfied. Next, D-flatness condition gives
\begin{align}
    \label{stringd-flat}
    f_1^2(r) - f_2^2(r) = 0 .
\end{align}
For the string solutions to satisfy eq. (\ref{stringd-flat}), we require $f_1(r) = f_2(r) \equiv f(r)$. Therefore the string solutions are
\begin{align}
    \label{x-stringsolm}
    h_1(x) = f(r) e^{im\theta} \left(\begin{array}{c} 0 \\ 1 \end{array}\right), \quad 
    h_2(x) = f(r) e^{-im\theta} \left(\begin{array}{c} 1 \\ 0 \end{array}\right), \quad 
    \vec{Z}(x) = -m\frac{z(r)}{r} \vec{e}_\theta .
\end{align}
On substituting eq. (\ref{x-stringsolm}) into the equations of motion on Lagrangian given in (\ref{susyXL}), they reduce to
\begin{align}
    \label{x-stringeom}
    \begin{aligned}
        &z'' - \frac{z'}{r} + 2m\alpha \left[\left(1 - \frac{\alpha}{2}z\right) f^2 \right] = 0 \\
    &{f''} + \frac{f'}{r} - m^2 \left(1 - \frac{\alpha}{2}z\right)^2 \frac{f}{r^2} -y^2 \left(f^2 - u^2 \right) f = 0 .
    \end{aligned}
\end{align}
where $\alpha \equiv \sqrt{g_1^2 + g_2^2}$. The equations in eq. (\ref{x-stringeom}) are the same as eq. (\ref{noeomf}) and eq. (\ref{noeomv}) by replacing $(f(r), z(r), \alpha, y^2, u)$ to $(f(r)/\sqrt{2}, a(r), 2e, 4\lambda, v/2)$. It indicates that $f(r)$ and $z(r)$ in this electroweak string are essentially equivalent to $f(r)$ and $a(r)$ in the Nielsen-Olesen string.

\section{Stability of non-topological string}

Because the first homotopy group of the Higgs vacuum is trivial, the string solutions given in eq. (\ref{x-stringsolm}) is not topologically stable. We shall investigate the stability of the string solutions by considering infinitesimal perturbations around them and finding whether the energy decreases or not. The energy functional for static and $z$-independent fields is denoted by $E = \int dz\, \mu$, where $\mu$ is 
\begin{align}
    \label{susyewmu}
    \mu = \int rdrd\theta \left[ \frac{1}{4} W^a_{\bar{i}\bar{j}} W^a_{\bar{i}\bar{j}} + \frac{1}{4} F_{\bar{i}\bar{j}} F_{\bar{i}\bar{j}} + \left| D_{\bar{i}} h_1 \right|^2 + \left| D_{\bar{i}} h_2 \right|^2 + \left|\partial_{\bar{i}} s \right|^2 + V(h_1, h_2, s) \right] .
\end{align}
Substituting eq. (\ref{x-stringsolm}) into eq. (\ref{susyewmu}), the string tension is obtained as
\begin{align}
    \label{x-stringmu}
    \mu = \int rdrd\theta \left[m^2 \frac{{z'}^2}{2r^2} + 2{f'}^2 + \frac{2m^2}{r^2} \left(1-\frac{\alpha}{2}z\right)^2 f^2 + y^2 \left( f^2 - u^2 \right)^2 \right] .
\end{align}
We shall set $m=1$ in the followings.

%\subsection{Analytical calculations for the variation of the energy linear density}
%In this subsection, w
%We consider  the perturbations which 
%are time independent and 
%the zero components of the gauge field to be zero, which can be realized by fixing gauge partially. The other gauge redundant degree will be fixed later. 

Before the detailed calculation, we note that we can ignore the $z$-dependent perturbations and the $z$-components of the vector fields.
%because of the $z$-independence of the string solutions (eq. (\ref{x-stringsolm})). 
The gradient energy terms, which are related with these perturbations,  are given by
\begin{align}
    \frac{1}{2} W^a_{\bar{i}z} W^a_{\bar{i}z} + \frac{1}{2} F_{\bar{i}z} F_{\bar{i}z} + \left| D_z h_1 \right|^2 + \left| D_z h_2 \right|^2 + \left|\partial_z s \right|^2 , 
\end{align}
which are separated from other variations and become positive. In addition, if the variation of potential energy made by the $z$-dependent perturbations is denoted as $\int dz\delta \mu(z)$, it always becomes larger than $\int dz \mu(z_{min})$ where $z_{min}$ minimizes the $\mu(z)$. For this reason, we ignore the $z$-dependent perturbations and the $z$-component of the gauge fields. Similar discussion is possible for $t$-dependent perturbations and $t$ component gauge fields. Thus, it is sufficient to check whether
the string tension decreases or not by perturbations instead of the string energy.
%en we disscuss the quadratic variations of the string tension to check the stability of the string solutions.

Because of unbroken $U(1)$ gauge symmetry, the quadratic variations separate to the neutral part $\mu_n$ and the charged part $\mu_c$ as in the calculation in section 2. The neutral part of the string tension is given by 
\begin{align}
    \label{xmun}
    \int rdrd\theta &\left[ \frac{1}{2} \left(\nabla\times\vec{Z}\right)^2 + \left| \left(\partial_{\bar{i}} + i\frac{\alpha}{2}Z_{\bar{i}}\right) \varphi_1 \right|^2 + \left| \left(\partial_{\bar{i}} - i\frac{\alpha}{2}Z_{\bar{i}}\right) \varphi_2 \right|^2 + \left| \partial_{\bar{i}} s \right|^2 + y^2 \left| \varphi_1 \varphi_2 - u^2 \right|^2 \right. \nonumber\\
    &\hspace{0.5cm} \left. + y^2 |s|^2 \left(\left| \varphi_1 \right|^2 + \left| \varphi_2 \right|^2 \right) + \frac{\alpha^2}{8} \left(\left| \varphi_1 \right|^2 - \left| \varphi_2 \right|^2 \right)^2 + \frac{1}{2} \left(\nabla\times\vec{A}\right)^2 \right] ,
\end{align}
where $\varphi_1$ and $\varphi_2$ are neutral components of $h_1$ and $h_2$, respectively. In eq. (\ref{xmun}), we denote the original string solution and the perturbations together as $\varphi_1, \varphi_2, s, Z_{\bar{i}}, A_{\bar{i}}$. The fourth, sixth and eighth terms are non negative and it can be zero by setting $s=0$ and $\nabla\times\vec{A} = 0$. Thus, we can ignore these terms. We have to check whether the perturbations of $\varphi_1, \varphi_2, Z_{\bar{i}}$ can make the string tension lower or not. However, this problem is equivalent to checking the stability of the Nielsen-Olesen string,
which appears when an $U(1)$ is spontaneously broken by developing the VEVs of two Higgs bosons. This has been discussed in Ref.\cite{La:1993je}. Because the Nielsen-Olesen string with two Higgs bosons is also a topological string,
% identically to the original Nielsen-Olesen string, 
it is stable and does not decay. Because the string tension (\ref{xmun}) with vanishing 
$s$ and $\nabla\times\vec{A}$ is nothing but the string tension of the Nielsen-Olesen string with two Higgs bosons, we conclude that the neutral perturbations does not reduce the string tension.

Let us consider whether the unbroken $U(1)$ charged perturbations destabilize the string solution or not. We redefine the Higgs fields $\phi_1$ and $\phi_2$ as
% To make a discussion more simple, we deform the two Higgs fields $h_1$ and $h_2$ as
\begin{align}
    \left(\begin{array}{c}
        \phi_1(x) \\ \phi_2(x)
    \end{array}\right)
    \equiv \frac{1}{\sqrt{2}}\left(\begin{array}{cc}
        1 & -1 \\ 1 & 1
    \end{array}\right)
    \left(\begin{array}{c}
        h_1(x) \\ -i \sigma_2 (h_2)^*(x)
    \end{array}\right) ,
\end{align}
where $\phi_1$ and $\phi_2$ has the same charge. On the $SU(2)\times U(1)\rightarrow U(1)$, $\phi_1$ develops a non-vanishing VEV, $\braket{\phi_1}=\sqrt{2}u$, while $\phi_2$ has vanishing  VEV. The Higgs potential can be rewritten as
%Deforming the potential in eq. (\ref{hpot}) by using $\phi_1$ and $\phi_2$, it is given as
\begin{align}
    \label{phipot}
    \tilde{V} (\phi_1, \phi_2) = &\frac{y^2}{4} \left( \left|\phi_1\right|^2 - \left|\phi_2\right|^2 - 2u^2\right)^2 +\frac{y^2}{4} \left| \phi_1^\dagger \phi_2 - \phi_2^\dagger \phi_1 \right|^2 + \frac{\alpha^2}{8} \left( \phi_1^\dagger \phi_2 + \phi_2^\dagger \phi_1 \right)^2 \nonumber\\
    & + \frac{g_2^2}{2} \left[ \left| \phi_1 \right|^2 \left| \phi_2 \right|^2 - (\phi_1^\dagger \phi_2) (\phi_2^\dagger \phi_1) \right] .
\end{align}
We consider the charged perturbations of $\phi_1$ and $\phi_2$ as
\begin{align}
    \label{phinotation}
    \phi_1(x) = \left(\begin{array}{c} \eta_1(x) \\ \sqrt{2} f(r) e^{i\theta} \end{array}\right), \quad \phi_2(x) = \left(\begin{array}{c} \eta_2(x) \\ 0 \end{array}\right).
\end{align}
%where $\eta_1$ and $\eta_2$ are the perturbations. 
Substituting eq. (\ref{phinotation}) to eq. (\ref{phipot}) and expanding the potential up to the second order of $\eta_1$ and $\eta_2$, we obtain
\begin{align}
    \label{phipotquadra}
    \tilde{V} (\phi_1, \phi_2) \sim y^2(u^2-f^2)^2- y^2 (u^2 - f^2) |\eta_1|^2 + y^2 (u^2 - f^2)|\eta_2|^2 + g_2^2f^2|\eta_2|^2.
\end{align}
In eq. (\ref{phipotquadra}), the terms which include $|\eta_2|^2$ give positive contributions to the potential.  In addition, the gradient energy for $\eta_2$ is positive because 
the kinetic term of $\phi_2$ at $\eta_2=0$ gives the minimum kinetic energy, 
$E_{\rm min}=0$.
 Therefore we conclude that $\eta_2$ does not destabilize the string solution.

The perturbations which may decrease the string tension are $\eta_1$ and the perturbations of $W_{\bar{i}}^{\bar{a}}$. However, these modes are nothing but the
dangerous modes in the electroweak string, which is discussed in section 2. 
%destabilize the string solution identically to the perturbations which destabilize the electroweak string. 
When we take $\phi_2=0$ on eq. (\ref{susyXL}), the Lagrangian becomes
the same as 
the Lagrangian of the electroweak string in eq. (\ref{ewtranslagrangian})
by corresponding the parameters $\lambda$ and $v$ to $y^2/4$ and $2u$,
respectively.
%\begin{align}
%    \tilde{V} (\phi_1, \phi_2=0) = \frac{\lambda^2}{4} \left( \phi_1^\dagger \phi_1 - 2v^2\right)^2 ,
%\end{align}
%which is nothing but the potential for the electroweak string 
%n the above equation, the shape of the potential of $\phi_1$ is the same as the potential which leads to the electroweak breaking. 
It indicates that the perturbations $\eta_1$ and $W_{\bar{i}}^{\bar{a}}$ for the string solution (\ref{x-stringsolm}) is equivalent to the charged perturbations of the electroweak string. Therefore we conclude that the stability of the string solution (\ref{x-stringsolm}) is the same as the stability of the electroweak string.

The parameter region in the parameter plane $(\sqrt{\beta}=m_H/m_Z,  \cos^2\theta_W)$ for the classically stable electroweak string is given in Ref.\cite{James:1992wb}. 
This parameter region can be 
applied to the parameter region for the classically stable non-topological string discussed in this paper. 

%The parameters which determine stability of the string solutions (\ref{x-stringsolm}) are $\beta ( \equiv \frac{2\lambda^2}{\alpha^2})$ and $\theta_W$ where $\tan\theta_W \equiv g_1/g_2$. They correspond to $\beta$ and $\theta_W$ in the discussion for the electroweak string. The string solution is classically stable only on the same region which the electroweak string is stable on.
%Detail of the parameter region refer to Ref. \cite{James:1992wb}.

\section{Discussion and summary}
We have extended the electroweak string  to the non-topological string in the SUSY $SU(2)\times U(1)$ gauge theory and have seen that its classical stability conditions are the same
as those for the electroweak string, with appropriate replacement of some parameters. Since the configuration of the electroweak string is embedded in the SUSY gauge theory, the most
dangerous mode which makes the electroweak string unstable appears also in the 
non-topological string in the SUSY gauge theory. 
Therefore, the classical stability conditions cannot
be weaker than those of the electroweak string. The question is whether the classical 
stability conditions become more severe or not. We have concluded that the classical 
stability conditions become the same as those in the electroweak string. We have shown
that the classical stability is determined only by two parameters, $(\cos\theta_W, \beta)$,
as in the electroweak string, and the other modes than the above dangerous mode do not destabilize the non-topological string configurations.

 The stability region in the $(\sin^2\theta_W,\sqrt{\beta})$ for the electroweak string has
been shown in Ref. \cite{James:1992zp, James:1992wb} by James, Perivolaropoulos, and Vachaspati. They have shown
that the electroweak string becomes classically stable only in the limited region where
$\cos^2\theta_W<0.1$ and $\beta\leq 1$. Therefore, not only the electroweak string in
the SM $(\sqrt{\beta}=1.4, \sin^2\theta_W=0.23)$ but also the non-topological string
in the SUSY $SU(2)_R\times U(1)_{B-L}$ gauge theory, which appears in $SO(10)$
grand unified theory $(\sin^2\theta_W=0.6)$, becomes classically unstable. 
 
%Although we have not found a well-motivated model beyond the SM in which 
%the stable non-topological string appears, 
We think that the study for the classical stability of 
non-topological string becomes important to test the models beyond the SM via 
gravitational wave detection. We hope that this work contributes to this interesting subject. 

\section*{acknowledgments}
The authors thank Minoru Eto for useful discussions. The authors also thank Rinku Maji and Yu Hamada for suggesting interesting papers. Y. K. thanks Motoko Fujiwara for the
discussion on the general two Higgs doublet model. 
This work is supported in part by the Grant-in-Aid for Scientific Research 
  from the Ministry of Education, Culture, Sports, 
 Science and Technology in Japan  No.~19K03823(N.M.).

\end{document}